\documentclass[conference]{IEEEtran}

\usepackage{booktabs} 

\usepackage{graphicx}
\usepackage{mdframed}
	\usepackage{amsmath}
	\usepackage{amsfonts}
    \usepackage{longtable}
	\usepackage{fancyhdr}
	\usepackage{mdframed}
\mdfdefinestyle{style1}{leftmargin=1cm,rightmargin=0.5cm,skipbelow=0.5cm,skipabove=0.2cm,
   innertopmargin=2pt}
	\usepackage{balance}
    \usepackage{fancybox}
    \usepackage{booktabs}
	\usepackage{enumerate}
    \usepackage{framed}
    \usepackage{pgfplots}
	\usepackage{listings}
    \usepackage{subcaption}
    \usepackage{xcolor}
    \usepackage{hyperref}

\ifCLASSOPTIONcompsoc
  \usepackage[nocompress]{cite}
\else
  \usepackage{cite}
\fi

\begin{document}
\title{Test Design and Review Argumentation in AI-Assisted Test Generation}

\author{
\IEEEauthorblockN{Eduard Paul Enoiu}
\IEEEauthorblockA{
\textit{Department of Computer Science and Engineering} \\
\textit{Mälardalen University} \\
Västerås, Sweden \\
eduard.paul.enoiu@mdu.se
}
\and
\IEEEauthorblockN{Robert Feldt}
\IEEEauthorblockA{
\textit{Chalmers University of Technology} \\
\textit{University of Gothenburg} \\
Gothenburg, Sweden \\
robert.feldt@chalmers.se
}
}

\IEEEtitleabstractindextext{%
\begin{abstract}

AI assistants can increasingly generate and evolve test cases. The challenge is no longer merely to produce them, but also to help engineers understand why a generated artefact exists and what supports it. Existing work has focused on classifying testing techniques, linking requirements to tests and structuring system assurance arguments, but it does not explicitly represent the argumentation behind individual test design decisions. We propose a conceptual taxonomy and a structured template for AI-assisted test generation that characterizes a test case by its test goal, claim, reason, and evidence. The taxonomy is intended for both constructive use during test design and retrospective use during review, to assess the quality of the attached argument rather than the plausibility or objective value of the generated test cases. 

\end{abstract}
}
%
%



\maketitle
\IEEEdisplaynontitleabstractindextext
\IEEEpeerreviewmaketitle

\section{Introduction}

AI assistants are increasingly used to propose, generate, and evolve test suites \cite{fakhoury2024llm,chen2025agenttester,harman2025mutation}, but the challenge is shifting from producing test cases to assuring their quality \cite{akbarova2026understanding,strandberg2025ethical,phillips2021four}. 
In this setting, trust aspects such as explainability \cite{jobin2019global,strandberg2025ethical} become an extra-functional requirement for AI-assisted test generation, enabling engineers to understand why a test case was created, what it supports, which assumptions it relies on, and why the results are credible. The motivation for this is quite pragmatic. Without such a rationale, generated test cases are difficult to assess, challenge, maintain, or trust, especially in review-intensive and safety-critical settings. 

Existing work provides only part of what is needed. The software testing literature \cite{ammann2016introduction,beer2008role} and standards \cite{srs} classify tests as goal-driven, context-dependent, and constrained. Requirement–test traceability \cite{barmi2011alignment} establishes which requirement a test addresses, and rationale-oriented software engineering \cite{tang2007rationale} supports reasoning more broadly. However, these lines of work do not explicitly articulate the rationale for individual test design decisions. Early evidence \cite{akbarova2026understanding} also suggests that LLMs can provide broad explanations during testing, particularly for boundary-value cases. 


In this paper, we propose the term \textit{test design argument} to refer to an inspectable rationale attached to a test case: the test goal it contributes to, the claim it is intended to support, the reason the test is needed in context, and the evidence a reviewer can inspect.  
For example, a boundary test may be created to support a claim about threshold behaviour because a requirement specifies a threshold; the requirement text, boundary rationale, and execution results then provide inspectable evidence. In this way, we address the missing middle ground between test design technique classification, traceability and assurance, and shift attention from the objective value of generated test cases alone to the quality of the associated argument.

\section{Related Work}

The ISO/IEC/IEEE 29119 \cite{2013i2s} family standardises test processes, documentation templates, and test design techniques. Similarly, the ISTQB \cite{spillner2021software} syllabus classifies techniques as specification-based, structure-based, and experience-based, with examples including equivalence partitioning, boundary value analysis, statement or branch testing, error guessing, and exploratory testing.

\begin{figure*}[!t]
\centering
\includegraphics[width=0.8\textwidth]{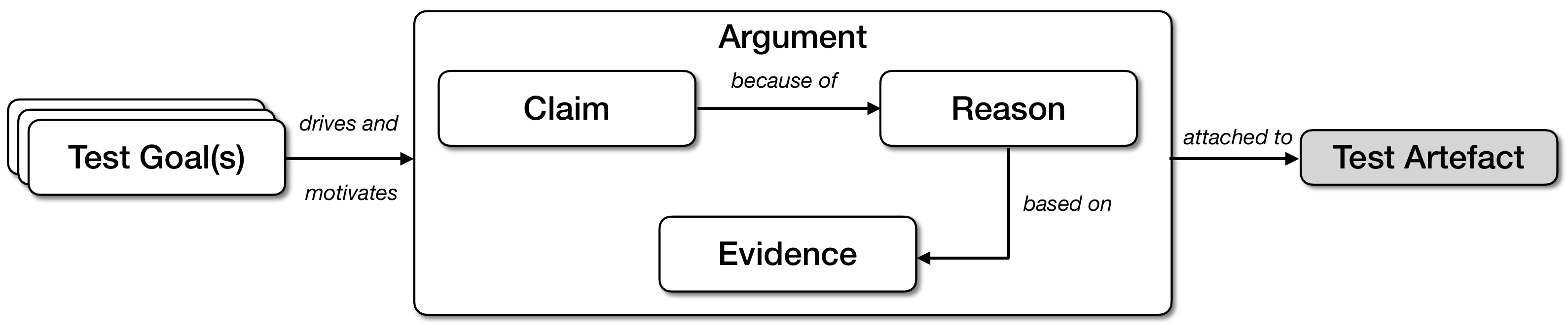}
\caption{Conceptual Model of a Test Design and Review Argument.}
\label{fig:tart} 
\end{figure*}
Regulated and safety-critical domains often require evidence that testing meets integrity and certification objectives. For example, ISO 26262-6 \cite{griessnig2017development} and EN 50128 \cite{en200150128} define software testing techniques and activities linked to safety integrity levels. In these contexts, safety cases and notations such as GSN \cite{spriggs2012gsn} structure claims, strategies, and supporting evidence.


Transparency and explainability are increasingly treated not merely as useful properties, but as trustworthiness-related extra-functional requirements for AI-enabled software engineering tools \cite{jobin2019global}. For AI-assisted test automation \cite{strandberg2025ethical}, this requirement translates into capabilities such as logging, monitoring, and traceable decision records, thereby making the assistant’s test actions inspectable. Explainability guidance highlights the need for meaningful, accurate, and clear explanations. NIST’s \cite{phillips2021four} four principles of Explainable AI (explanation, meaningfulness, explanation accuracy, and knowledge limits) highlight transparency, documentation, and risk management, which are directly relevant when AI systems participate in test design and must justify proposed test cases. 


Itkonen et al. \cite{itkonen2009testers} identified 22 test practices, showing that test design is a problem-solving activity that blends exploration and documentation-driven work, which motivates our view that test design should not only classify how tests are derived. Work on testers’ cognitive processes \cite{enoiu2020towards} explicitly argues for modelling test design as cognitive problem-solving, aligning with the framing that test design rationale should reflect human reasoning and evidence argumentation.

Previous research on requirement–test alignment and traceability \cite{barmi2011alignment} helps establish which requirement a test relates to, but it does not make the argumentative structure explicit. More broadly, prior work in software engineering \cite{tang2007rationale,burge2008rationale} has long emphasized the capture of rationale to support traceability and reasoning, particularly for design and architectural decisions.

\begin{mdframed}[linewidth=1pt, linecolor=black, backgroundcolor=gray!10, roundcorner=5pt]
\small{\textbf{Gap:} Existing work classifies testing approaches, recovers trace links among software artefacts\footnote{By software artefacts we mean the software work products produced and used across software engineering activities.}, and structures assurance arguments at the system level, but not the argumentation behind individual test design decisions. We address this gap by linking each generated test to its reason, claim, argument, and evidence.}
\end{mdframed}

\section{Test Design Argument and Reasons Taxonomy}

This section proposes a conceptual taxonomy of test design arguments and reasons intended for test engineering work and for AI-assisted workflows. The design is explicitly aligned with established test objectives and technique taxonomies, as well as assurance case claim, argument, and evidence patterns.
As illustrated in Figure \ref{fig:tart}, the proposed conceptual model treats a test artefact as linked to test goal(s) through a structured argument, in which reasons support claims and evidence substantiates those reasons.

\subsection{Conceptual Foundation: Arguments Attached to Test Cases}

The idea of attaching arguments to test cases can be understood in terms of Booth et al.'s model of research argumentation \cite{booth2009craft}. Just as an argument links a claim to supporting reasons and evidence, we treat each test case as carrying a compact justification for why it supports a test engineering claim. A test is therefore not only an artefact, but also an inspectable argument that makes its claim, assumptions, context, evidence, and limits explicit. In this framing, the attached argument makes explicit the warrant linking reason and claim, showing why a particular test case, under stated assumptions and evidence, supports a broader test goal.
Therefore, following the claim, reason and evidence model, we propose the following definition:
\begin{mdframed}[linewidth=1pt, linecolor=black, backgroundcolor=gray!10, roundcorner=5pt]
\small{\textbf{Definition:} A \textit{Test Design Argument} is a structured statement asserting that a test artefact is created to support a claim because of a reason, based on evidence, and that this argument contributes to a test goal.}
\end{mdframed}
To clarify these terms, a \textit{reason} explains why a test artefact was created in context, a \textit{claim} states what the test case is intended to demonstrate, an \textit{argument} represents the overall justification linking reason, claim and evidence, and \textit{evidence} refers to the inspectable artefacts used to assess that justification.

\subsection{From Test Goals to Arguments}
We organize this part of the taxonomy around the relationship between a \textit{test goal}, a \textit{test artefact}, and the \textit{argument} that connects them. The test goal records the broader testing objective \cite{enoiu2023understanding}, while the test artefact captures the concrete test case and the test specification created. The argument explains why that artefact exists and why it is relevant in context. It includes three parts: the \textit{reason}, which states why this test case is needed in context,
 the \textit{claim}, which states what the test case is intended to demonstrate with respect to the test goal; and the \textit{evidence}, which consists of inspectable artefacts that allow a reviewer to evaluate whether the reason supports the claim. 

This separation matters because test artefacts may support the same goal for different reasons, claims, and evidence, while similar reasons may lead to different claims. Reasons also vary in their level of structure, ranging from semi-formal design rationale to judgments and suspicions about weak areas. Treating reasons as a continuum reflects how test design moves between exploratory intent and evidence-based argumentation.
This framing also aligns with observed manual testing practices. Itkonen et al. \cite{itkonen2009testers,itkonen2012role} show that strategies and techniques map to different test goals, such as boundary testing, input combinations, and exploring weak areas. In our framing, these practices describe not only how test cases are created, but also the basis for the argument linking a test artefact to its broader goal. In safety-critical settings, some measures provide more well-defined grounds for such arguments, while others, such as error guessing, are more experience-based. Making this explicit supports inspectable and reviewable test rationale for both humans and AI assistants.
This is crucial for mixed human–AI environments, because AI can often generate plausible but unfounded rationales \cite{fan2023large,jobin2019global}. Therefore, capturing such reasons could provide the additional information needed for review and accountability.

\subsection{Claim and Reason Patterns}
The conceptual taxonomy treats the claim as an explicit statement of what a test artefact is intended to demonstrate or provide evidence for. The claim is distinct from both the reason for creating the test and the evidence used to justify it. It instead expresses the intended contribution of the test case and its intent \cite{booth2009craft}. The claim patterns are synthesized from established testing objectives and techniques \cite{ammann2016introduction,spillner2021software}, including \textit{requirement satisfaction claims, fault revealing claims, coverage contribution claims, risk mitigation claims, regression-preservation claims and operational confidence claims}. 

While not shown explicitly in Figure \ref{fig:tart}, each claim depends on a warrant\footnote{A warrant is not the whole argument. It is the underlying principle or logic that explains why this kind of reason can support this kind of claim.}, that is, an underlying line of reasoning explaining why a reason is used to support a claim (e.g., \textit{cause and effect, certain rules, a definition, or a principle of reasoning}). The \textit{reason} explains why the test case is needed in its specific context and captures the specific motivation for creating the test artefact. A reason may be based on the specific interpretation of a text, analogy or precedent, reasoning tied to the test goal, customary testing practice, convention, or standards. This classification is adapted from broader traditions in legal interpretation and reasoning and applied here to the justification of claims \cite{sep-legal-interpretation}. 

\subsection{Evidence Classes}
Evidence in our taxonomy is the information used to determine whether a reason supports the claim for a given test case, and thus supports the test design argument.
To make the taxonomy usable for justification, evidence can be modeled explicitly with quality aspects and evidence types: documents, test results, logs, traces, screenshots and oracles used, coverage artefacts and metrics (e.g., branch, MC/DC, interaction coverage), risks, performance measurements, defect history, reproduction scripts, and static analysis outputs. This list is intentionally non-exhaustive, as the relevance of evidence depends on the system, domain, and testing goals. Aligned with ISTQB \cite{spillner2021software}, 
such evidence may be grouped into three broad classes: \textit{test basis and other work products, such as requirements, user stories, designs, and code; execution results and their oracles; and traceability and coverage artefacts} used to evaluate adequacy.

Beyond identifying evidence types, the taxonomy should also capture quality aspects such as relevance to the claim. For example, vague references to coverage provide weaker support than evidence explicitly linked to a coverage specification and execution artefacts. Evidence is also stronger when it can be independently inspected, repeated, or traced to a requirement. Making this explicit helps distinguish just plausible AI-generated justifications from genuinely reviewable support for a test design argument.

\section{How Humans and AI Assistants can use the Taxonomy during Test Design}
This conceptual taxonomy is intended for two complementary modes of use: constructive use, during test design and generation, and retrospective use, during review. In constructive use, the taxonomy helps a human engineer or an AI assistant make the rationale for a test case explicit as it is designed. A constructive workflow may proceed as follows: \textit{identify the test goal, record the rationale, select or formulate the claim, explain why the test result supports that claim, create the executable test artefact, and attach evidence}. In retrospective use, the taxonomy helps reviewers assess whether the rationale is inspectable and supported. In this mode, the reviewer checks whether the test case includes: \textit{a clear reason, an explicit test goal, a specific claim, a plausible argument, and inspectable evidence}. 
We instantiate the taxonomy as a structured argument template. 
\begin{mdframed}[linewidth=1pt, linecolor=black, backgroundcolor=yellow!10, roundcorner=5pt]
\textbf{Structured Argument Template:}
\small{
\begin{enumerate}
    \item \textbf{Test Goal [G]:} Which test goal does test case \textit{X} support?
    \item \textbf{Reason [R]:} Why is \textit{X} needed in this context?
    \item \textbf{Claim [C]:} What is \textit{X} intended to demonstrate?
    \item \textbf{Evidence [E]:} What artefacts support the argument?
    \item \textbf{Argument:} Test Case \textit{X} is created to support claim \textit{[C]}
    because of reason \textit{[R]}, based on evidence \textit{[E]}, and this contributes
    to test goal \textit{[G]}.
\end{enumerate}}
\end{mdframed}
This template can be used either by a human engineer to document the rationale for a test case or by an AI assistant when explicitly prompted to generate that rationale. 

To showcase this template, we use a trip-logic program in a safety-critical reactor-protection software system used in a previous study by Shin et al. \cite{shin2012empirical}. We first illustrate the structured argument with a relatively well-defined case in which the reason is derived based on the specification and the supporting evidence is straightforward. In this example, the trip signal is triggered when the monitored signals fall outside a safe interval or when fault-related input conditions occur.

\begin{mdframed}[linewidth=1pt, linecolor=black, backgroundcolor=yellow!10, roundcorner=5pt]
\footnotesize{
\textbf{Test Case:} $t\_low$ (Inputs: $f\_X = -56$, $f\_Module\_Error = false$, $f\_Channel\_Error = false$, and $th\_X\_Logic\_Trip = false$; Expected Output: $th\_X\_Trip = true$). 
\newline

\noindent \textbf{[G]:} Requirement satisfaction and fault detection related to boundaries.
\newline
\textbf{[C]:} Test Case $t\_low$ supports the claim that the controller activates the trip signal when the temperature is below the threshold. 
\newline
\textbf{[R]:} 
Because $t\_low$ exercises the safety-critical lower threshold and should be tested at the boundary.
\newline
\textbf{[E]:} The requirement specification ID, test input and output specification, and execution logs.}

\end{mdframed}

With this argument, a reviewer can inspect why this test exists, what claim it is meant to support, and what evidence makes that claim reviewable.

By contrast, a less well-defined case may involve multiple fault signals near the safety-range boundary, where the specification is unclear about their interactions. In such a case, the reason may come from practitioner experience and prior defect patterns in related systems; the claim becomes more exploratory than requirement-derived, and the evidence is partial rather than straightforward. The following partial example illustrates that the taxonomy is useful not only for well-defined cases but also for cases in which reasons, claims, and evidence are weaker and must be made explicit for review:

\begin{mdframed}[linewidth=1pt, linecolor=black, backgroundcolor=yellow!10, roundcorner=5pt]
\footnotesize{
\textbf{Test Case:} $t\_fault$ (Inputs: $f\_X = -54$, $f\_Module\_Error = true$, $f\_Channel\_Error = true$, and $th\_X\_Logic\_Trip = false$). 
\newline

\noindent \textbf{[G]:} Exploratory fault detection related to boundary fault interactions.
\newline
\textbf{[C]:} Test Case $t\_fault$ explores whether the trip signal behaves correctly when multiple fault flags occur near the safe-range boundary.
\newline
\textbf{[R]:} Experience from a related project suggests that when multiple fault conditions are active simultaneously, the trip logic may behave unexpectedly near boundary values. 
\newline
\textbf{[E]:} Defect history from a related project, absence of specification for this input combination, and the execution logs.
}
\end{mdframed}

Without this explicit argument, $t\_fault$ could appear to be an arbitrary input combination; with this information, a reviewer can see that it is an exploratory test motivated by prior defect history and a specification gap.

An AI assistant can use the taxonomy to make test-design reasoning explicit, rather than merely generating test cases.
The taxonomy can also be used to assess AI assistants based on the quality of the argument associated with a generated test case, rather than solely on the artefact's plausibility. In an AI-assisted workflow, the assistant could generate a test case along with the corresponding [G], [R], [C], and [E] fields, which would then be checked and reviewed. While this does not guarantee correctness, it may make test design argumentation more explicit and easier to scrutinize.

\section{Conclusion}
The conceptual taxonomy explains why a test case was created by making explicit its reason, test goal, claim, and evidence. Each specific argument implies a different kind of justification and expected evidence, making test intent explicit and inspectable for both humans and AI-assisted test generation. This taxonomy is intended both as a test-design aid for AI-assisted test generation and as an evaluation lens for assessing the quality of generated test arguments.
This work has several limitations. The taxonomy is conceptual and non-exhaustive and should be refined through a more systematic development and validation process. We have not evaluated how it can be applied to assess the outputs of different AI assistants. Moreover, arguments are not necessarily correct or useful. Future work should refine the taxonomy, validate it with practitioners, compare human- and AI-generated test arguments, and develop metrics and tools for assessing argument and evidence quality.

\section{Acknowledgement}
This work was supported by Software Center project 68 (TRACE), Vinnova-funded ITEA projects MONA LISA and MATISSE (101140216), and the AI and Society Fellowship. Robert Feldt was supported by Wallenberg AI, Autonomous, and Software Program (WASP, BoundMiner project) and the Chalmer's Foundation's Academic Excellence Program.

\balance
\bibliographystyle{IEEEtran}
\bibliography{acmart} 

\begin{thebibliography}{10}
\providecommand{\url}[1]{#1}
\csname url@samestyle\endcsname
\providecommand{\newblock}{\relax}
\providecommand{\bibinfo}[2]{#2}
\providecommand{\BIBentrySTDinterwordspacing}{\spaceskip=0pt\relax}
\providecommand{\BIBentryALTinterwordstretchfactor}{4}
\providecommand{\BIBentryALTinterwordspacing}{\spaceskip=\fontdimen2\font plus
\BIBentryALTinterwordstretchfactor\fontdimen3\font minus \fontdimen4\font\relax}
\providecommand{\BIBforeignlanguage}[2]{{%
\expandafter\ifx\csname l@#1\endcsname\relax
\typeout{** WARNING: IEEEtran.bst: No hyphenation pattern has been}%
\typeout{** loaded for the language `#1'. Using the pattern for}%
\typeout{** the default language instead.}%
\else
\language=\csname l@#1\endcsname
\fi
#2}}
\providecommand{\BIBdecl}{\relax}
\BIBdecl

\bibitem{fakhoury2024llm}
S.~Fakhoury, A.~Naik, G.~Sakkas, S.~Chakraborty, and S.~K. Lahiri, ``{LLM}-based test-driven interactive code generation: User study and empirical evaluation,'' \emph{Transactions on Software Engineering}, vol.~50, no.~9, pp. 2254--2268, 2024.

\bibitem{chen2025agenttester}
H.~Chen, K.~Chen, F.~Zhang, T.~Wang, and L.~Cheng, ``{AgentTester}: An {LLM}-based tool for unit test generation with automatically generated prompts,'' in \emph{International Conference on Intelligent Computing}.\hskip 1em plus 0.5em minus 0.4em\relax Springer, 2025, pp. 114--126.

\bibitem{harman2025mutation}
M.~Harman, J.~Ritchey, I.~Harper, S.~Sengupta, K.~Mao, A.~Gulati, C.~Foster, and H.~Robert, ``Mutation-guided {LLM}-based test generation at meta,'' in \emph{International Conference on the Foundations of Software Engineering}.\hskip 1em plus 0.5em minus 0.4em\relax ACM, 2025, pp. 180--191.

\bibitem{akbarova2026understanding}
S.~Akbarova, F.~Dobslaw, and R.~Feldt, ``Understanding on the edge: {LLM}-generated boundary test explanations,'' \emph{arXiv preprint arXiv:2601.22791}, 2026.

\bibitem{strandberg2025ethical}
P.~E. Strandberg, E.~P. Enoiu, and M.~Frasheri, ``Ethical challenges and software test automation,'' \emph{AI and Ethics}, vol.~5, no.~6, pp. 6185--6206, 2025.

\bibitem{phillips2021four}
P.~J. Phillips, C.~A. Hahn, P.~C. Fontana, A.~N. Yates, K.~Greene, D.~A. Broniatowski, and M.~A. Przybocki, ``Four principles of explainable artificial intelligence,'' \emph{NISTIR 8312 Report}, 2021.

\bibitem{jobin2019global}
A.~Jobin, M.~Ienca, and E.~Vayena, ``The global landscape of {AI} ethics guidelines,'' \emph{Nature Machine Intelligence}, vol.~1, no.~9, pp. 389--399, 2019.

\bibitem{ammann2016introduction}
P.~Ammann and J.~Offutt, \emph{Introduction to software testing}.\hskip 1em plus 0.5em minus 0.4em\relax Cambridge University Press, 2016.

\bibitem{beer2008role}
A.~Beer and R.~Ramler, ``The role of experience in software testing practice,'' in \emph{Euromicro Conference Software Engineering and Advanced Applications}.\hskip 1em plus 0.5em minus 0.4em\relax IEEE, 2008, pp. 258--265.

\bibitem{srs}
\emph{\BIBforeignlanguage{eng}{IEEE Standard for Software Quality Assurance Plans (730-1981)}}.\hskip 1em plus 0.5em minus 0.4em\relax USA: IEEE, 1981-11-13.

\bibitem{barmi2011alignment}
Z.~A. Barmi, A.~H. Ebrahimi, and R.~Feldt, ``Alignment of requirements specification and testing: A systematic mapping study,'' in \emph{International Conference on Software Testing, Verification and Validation Workshops}.\hskip 1em plus 0.5em minus 0.4em\relax IEEE, 2011, pp. 476--485.

\bibitem{tang2007rationale}
A.~Tang, Y.~Jin, and J.~Han, ``A rationale-based architecture model for design traceability and reasoning,'' \emph{Journal of Systems and Software}, vol.~80, no.~6, pp. 918--934, 2007.

\bibitem{2013i2s}
``Iso/iec/ieee 29119-1:2013(e): Software and systems engineering software testing part 1:concepts and [elektronisk resurs]...'' 2013.

\bibitem{spillner2021software}
A.~Spillner and T.~Linz, \emph{Software testing foundations: A study guide for the certified tester exam-foundation level-ISTQB{\textregistered} compliant}.\hskip 1em plus 0.5em minus 0.4em\relax dpunkt. verlag, 2021.

\bibitem{griessnig2017development}
G.~Griessnig and A.~Schnellbach, ``Development of the 2nd edition of the iso 26262,'' in \emph{European Conference on Software Process Improvement}.\hskip 1em plus 0.5em minus 0.4em\relax Springer, 2017, pp. 535--546.

\bibitem{en200150128}
CENELEC, ``{50128: Railway Application--Communications, Signaling and Processing Systems--Software for Railway Control and Protection Systems},'' in \emph{Standard Report}, 2001.

\bibitem{spriggs2012gsn}
J.~Spriggs, \emph{GSN-the goal structuring notation: A structured approach to presenting arguments}.\hskip 1em plus 0.5em minus 0.4em\relax Springer Science \& Business Media, 2012.

\bibitem{itkonen2009testers}
J.~Itkonen, M.~V. Mantyla, and C.~Lassenius, ``How do testers do it? an exploratory study on manual testing practices,'' in \emph{International Symposium on Empirical Software Engineering and Measurement}.\hskip 1em plus 0.5em minus 0.4em\relax IEEE, 2009, pp. 494--497.

\bibitem{enoiu2020towards}
E.~Enoiu, G.~Tukseferi, and R.~Feldt, ``Towards a model of testers' cognitive processes: Software testing as a problem solving approach,'' in \emph{QRS}.\hskip 1em plus 0.5em minus 0.4em\relax IEEE, 2020, pp. 272--279.

\bibitem{burge2008rationale}
J.~E. Burge, J.~M. Carroll, R.~McCall, and I.~Mistrik, \emph{Rationale-based software engineering}.\hskip 1em plus 0.5em minus 0.4em\relax Springer, 2008.

\bibitem{booth2009craft}
W.~C. Booth, G.~G. Colomb, and J.~M. Williams, \emph{The craft of research}.\hskip 1em plus 0.5em minus 0.4em\relax University of Chicago press, 2009.

\bibitem{enoiu2023understanding}
E.~P. Enoiu, G.~Gay, J.~Esber, and R.~Feldt, ``Understanding problem solving in software testing: An exploration of tester routines and behavior,'' in \emph{IFIP International Conference on Testing Software and Systems}.\hskip 1em plus 0.5em minus 0.4em\relax Springer, 2023, pp. 143--159.

\bibitem{itkonen2012role}
J.~Itkonen, M.~V. M{\"a}ntyl{\"a}, and C.~Lassenius, ``The role of the tester's knowledge in exploratory software testing,'' \emph{Transactions on Software Engineering}, vol.~39, no.~5, pp. 707--724, 2012.

\bibitem{fan2023large}
A.~Fan, B.~Gokkaya, M.~Harman, M.~Lyubarskiy, S.~Sengupta, S.~Yoo, and J.~M. Zhang, ``Large language models for software engineering: Survey and open problems,'' in \emph{International Conference on Software Engineering: Future of Software Engineering}.\hskip 1em plus 0.5em minus 0.4em\relax IEEE, 2023, pp. 31--53.

\bibitem{sep-legal-interpretation}
M.~Greenberg, ``{Legal Interpretation},'' in \emph{The {Stanford} Encyclopedia of Philosophy}, {F}all 2021~ed., E.~N. Zalta, Ed.\hskip 1em plus 0.5em minus 0.4em\relax Metaphysics Research Lab, Stanford University, 2021.

\bibitem{shin2012empirical}
D.~Shin, E.~Jee, and D.-H. Bae, ``Empirical evaluation on fbd model-based test coverage criteria using mutation analysis,'' in \emph{International Conference on Model Driven Engineering Languages and Systems}.\hskip 1em plus 0.5em minus 0.4em\relax Springer, 2012, pp. 465--479.

\end{thebibliography}

\end{document}